# A Join Index for XML Data Warehouses

Hadj Mahboubi, Kamel Aouiche and Jérôme Darmont

University of Lyon (ERIC Lyon 2)
5 avenue Pierre Mendès-France
69676 Bron Cedex
France
{first-name.last-name}@eric.univ-lyon2.fr

## *Abstract*
XML data warehouses form an interesting basis for decision-support applications that exploit complex data. However, native-XML database management systems (DBMSs) currently bear limited performances and it is necessary to research for ways to optimize them. In this paper, we propose a new join index that is specifically adapted to the multidimensional architecture of XML warehouses. It eliminates join operations while preserving the information contained in the original warehouse. A theoretical study and experimental results demonstrate the efficiency of our join index. They also show that native XML DBMSs can compete with XML-compatible, relational DBMSs when warehousing and analyzing XML data.

## 1. Introduction

Decision-support applications nowadays exploit heterogeneous data from various sources. Furthermore, the development of the Web 2.0 and the proliferation of multimedia documents contributed to the analysis of data are not only numerical or symbolic. These so-called *complex data* may indeed fall in several of the following categories (Darmont et al. 2005): data represented in various formats (databases, texts, images, sounds, videos...); diversely structured data (relational databases, XML document repositories...); data originating from several different sources (distributed databases, the Web...); data described through several channels or points of view (radiographies and audio diagnosis of a physician, data expressed in different scales or languages...); data that change in terms of definition or value (temporal databases, periodical surveys...). For example, analyzing medical data may lead to exploit jointly information under various forms: patient records (classical database), medical history (text), radiographies, echographs (multimedia documents), physician diagnoses (texts or audio recordings), etc.

In this context, XML proves a very interesting tool in the process of integrating and warehousing complex data for analysis (Darmont et al. 2003). However, decision-support queries are generally complex because they involve several join and aggregation operations. In addition, many native XML database management systems (DBMSs) present relatively poor performances when data volume is very large and queries are complex. Thus, it is crucial to design XML data warehouses that guarantee the best performance when accessing data. Indexing is one of the most frequently used techniques to achieve this goal.

Several solutions have been proposed for XML data indexing in the literature. However, the existing techniques support single labeled path expressions within one XML document. A path expression helps explore an XML document and extract a specific node (element) or sub-tree (sub-document). It cannot perform a join operation over several XML documents. In the context of XML data warehouses, decision-support queries are complex and involve several path expressions. Data are also generally distributed into several XML documents due to their large volume. Hence, XML queries should use specific indices to navigate these documents.

In this paper, we propose a new join index structure that is specifically adapted to multidimensional XML data warehouses. This structure is able to maintain a star schema of several XML documents and to preserve the information contained in these documents. It is actually a join index that ensures a faster execution of XQuery decision-support queries by eliminating join costs.

We theoretically and experimentally demonstrate that the use of our index significantly reduces the execution time of XQuery decision-support queries expressed on a data warehouse. In addition, our experiments show that, in our context, XML native DBMSs are competitive with relational, XML-compatible DBMSs.

The remainder of this paper is organized as follows. We first present the state of the art on XML indexing in Section 2. Then, we outline the context of this study in Section 3. We detail our join index structure in Section 4. In order to validate our proposal, we present a theoretical study in Section 5 and some experimental results in Section 6. Finally, we conclude the paper and discuss research perspectives in Section 7.

## 2. Related work

In this section, we assume that an XML document is defined as a labeled graph whose nodes represent document elements or attributes, and edges represent the element-subelement (or parent-child) relationship. Edges are labeled with element or attribute names.

Several studies address the issue of XML data indexing (Goldman & Widom 1997; Milo & Suciu, 1999; Cooper et al. 2001; Chung et al. 2002). They are more particularly devoted to optimize XML path expressions. Generally, they propose the creation of a schema or model (data structure) for XML data indexing. In practice, the index is a new XML document that is accessed instead of the original document.

The data guide is a summary structure for semi-structured and XML data (Goldman & Widom 1997). This index' structure describes by one single label all the nodes whose labels (names) are identical. The definition of the data guide is based on targeted path sets, i.e., the nodes that are reached by traversing this path.

The 1-index clusters nodes that share the same path in the XML data graph (Milo & Suciu 1999). This process is performed through a bi-simulation relationship. A 1-index is normally smaller than the initial data graph and thereby facilitates query evaluation. To operate a selection of labels or a path expression, a hash table or a B-tree structure is used to index graph labels.

The data guide and the 1-index code all the paths from the root node. The size of such summary structures or indices may be large, even larger than the original XML document when XML data are represented as graphs (cyclic XML document). This problem causes degradation in query performance. The A(k)-index, a variant of the 1-index, was proposed to overcome these shortcomings (Kaushik et al. 2002). It is based on the notion of k-dissimilarity and builds an approximate index to reduce the size of the index graph. An A(k)-index can retrieve, without referring to the data graph, path expressions of length of at most k, where k controls the resolution of the index and influences its size in a proportional manner. For large values of k, index size may become very large and this approach suffers from the same issues than the 1-index. For small values of k, the size of the index is substantially smaller, but it cannot handle long path expressions.

To accommodate path expressions of various lengths, without unnecessarily increasing the size of the whole index, the D(k)-index assigns different values of k to different index nodes (Qun et al. 2003). These values conform to a given set of frequently-used path expressions (FUPs). Large values of k are assigned to parts of the index corresponding to parts of the data graph targeted by long path expressions; while small values of k are assigned to parts of the index corresponding to data targeted by short path expressions. To facilitate the evaluation of path expressions with branching, a variant called the UD(k,l)-index also imposes downward similarity (Wu et al. 2003).

The AD(k)-index builds a coarser index than the A(k)-index, but suffers from a problem of over-refinement (He & Yang 2003). The M(k)-index, an improvement of the D(k)-index, solves the problem of large scan space in the index, without affecting path coverage. The drawback of its design, though, lies in the requirement to adapt to a given list of FUPs.

The U(*)-index (universal, generic index; Boulos & Karakashian 2006) also exploits the notion of bisimilarity (like the 1-index). However, to overcome the problem of large search space for XPath evaluations, the index structure uses a special node labeling scheme that enables pruning the search space. Furthermore, the U(*)-index does not need to be adapted to any particular list of FUP; it has a uniform resolution, and hence is more generic.

APEX is an adaptive index that searches for a trade-off between size and effectiveness (Chang et al. 2002). Instead of indexing all the paths from the root, APEX only indexes the FUPs and preserves the source data structure in a tree. However, since FUPs are stored in the index, path query processing is quite efficient. APEX is also workload-aware, i.e., it can be dynamically updated according to changes in the query workload. A data mining method is used to extract FUPs from the workload for incremental update (Agrawal & Srikant 1995).

Unfortunately, all these indexing techniques are ill-suited to decision-support queries. Data structures such as the data guide, the 1-index and its variants, and APEX are indeed applicable only on XML data that are targeted by simple path expressions. However, in the context of XML data warehouses, queries are complex and include several path expressions that compute join operations. Moreover, these indices operate on one XML document only, whereas in XML warehouses, data are managed in several XML documents and decision-support queries are performed over these documents.

Finally, other techniques such as the extended inverted list (Zhang et al. 2001) and Fabric (Cooper et al. 2001) are aimed at processing containment queries over XML data stored in relational databases. The extended inverted list includes a text index (T-index; Milo & Suciu 1999) that is similar to the traditional indices in information retrieval systems, and an element index (E-index) that maps elements into inverted lists. Fabric indexes several XML documents by encoding paths, from the root to the leaf nodes. These operations are performed by indicators that code path labels. These codes are then inserted in a Patricia trie (Cooper et al. 2001), which processes them like simple characters. However, Fabric is not adapted to XML data warehouses either, because it does not take into account the relationships that exist between XML documents in a warehouse (facts and dimensions). This index is thus not beneficial to decision-support queries.

For more details about XML indexing, the interested reader may refer to a recent, complete survey (Gou & Chirkova 2007).

# 3. Study context

## 3.1. XML data warehouse specification

Several studies address the issue of designing and building XML data warehouses. They propose to use XML documents to manage or represent facts and dimensions. The main objective of these approaches is to enable a native storage of the warehouse and its easy interrogation with XML query languages.

For instance, Pokornÿ (2003) models a star schema in XML by defining dimension hierarchies as sets of logically connected collections of XML data, and facts as XML data elements. Park *et al.* (2005) propose an XML multidimensional model in which each fact is described by a single XML document and dimension data are grouped into a repository of XML documents. Rusu *et al.* (2005) build facts and dimensions from XML documents generated through XQueries. Eventually, Hümmer *et al.* (2003) propose a family of templates, called XCube, to describe a multidimensional structure (dimension and fact data) for integrating several data warehouses into a virtual or federated data warehouse. All these approaches assume that the warehouse is composed of XML documents that represent both facts and dimensions. They are used when dimensions are dynamic and allow the support of end-user analytical tools.

All these studies more or less converge toward a unified XML warehouse model. They mostly differ in the way dimensions are handled and the number of XML documents that are used to store facts and dimensions. To illustrate the application of our indexing strategy, we selected in this paper the XCube specification to model a reference XML data warehouse. Since other models from the literature are quite similar, this is not a binding choice.

The advantage of XCube is its simple structure for representing facts and dimensions in a star schema. One XML document is used to represent dimensions and another one to represent facts. Hence, our reference data warehouse is composed of the following XML documents: *Schema.xml* specifies the data warehouse metadata; *Dimensions.xml* defines all the dimensions, characterized by their attributes and their values; and *Facts.xml* specifies the facts, i.e., dimension identifiers and measure descriptions and values.

The *Facts.xml* document stores facts. Its tree structure is described in Figure 1(a). The document root node, *CubeFact*, has one child, *cube*, which is itself composed of *cell* nodes defining facts: *fact* nodes (measures) and dimension references. A *fact* node has two attributes, *@id* and *@value*, which define the measure name and value, respectively. The

*dimension* node has two attributes, *@id* and *@value,* which define the dimensions name and its identifier's value, respectively.

The *Dimension.xml* document stores dimensions. Its tree structure is described in Figure 1(b). The document root node, *dimensionData*, has one child, *classification*, which is itself composed of *Level* nodes. A *Level* node is composed of *node* nodes defining dimension instances. A *node* is composed of *attribute* nodes that define the attributes of a dimension and their values (*@name* and *@value*).

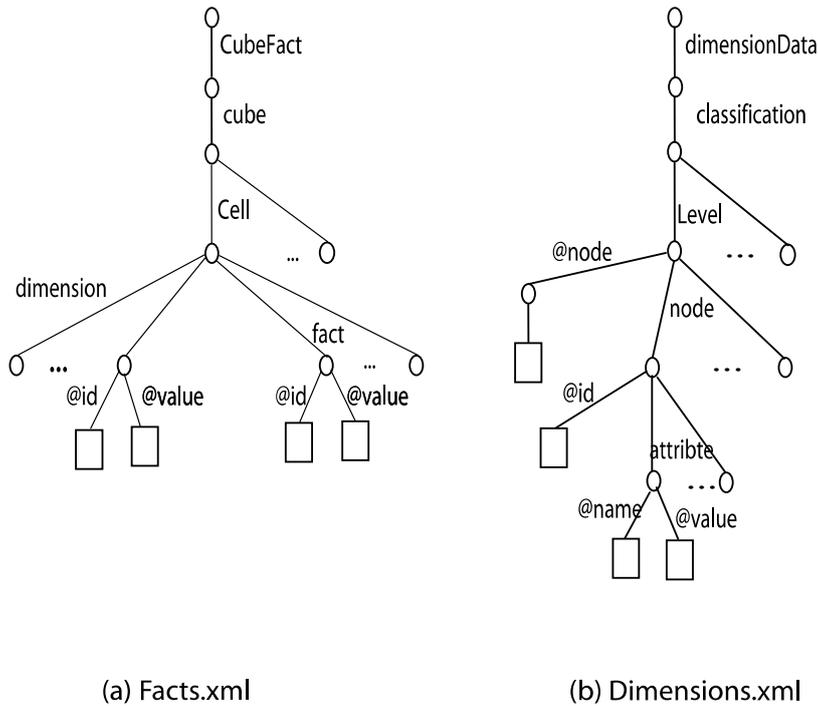

(a) Facts.xml     (b) Dimensions.xml

**Figure 1:** Dimension and fact document's structure

## 3.2. XML data warehouse interrogation

We selected the XQuery language (Boag et al. 2004) to formulate decision-support queries because, unlike simpler languages such as XPath, it allows complex queries, including join queries over multiple XML documents, to be expressed with the FLWOR syntax.

However, the XQuery language does not support the type of queries that are common in business analysis (Beyer et al., 2005). XQuery does indeed not include an explicit grouping construct comparable to the *group by* clause in SQL. Hence, several papers propose to extend XQuery to formulate decision-oriented queries (Borkar & Carey 2004; Beyer et al. 2005). In our implementation, we acknowledge this effort by adding to FLWOR expressions explicit *group by* clauses. More precisely, we added two functions: *group by (attribute list)* and *aggregation (aggregation operations, measure list)*, to the XQuery syntax. Figure 2 provides an example of decision-support query with a multiple *group by* clause that exploits these functions.

```
for $a in //dimensionData/classification/Level[@node='customers']/node,
$x in //CubeFacts/cube/Cell
let $q := $b/attribute[@name='cust name']/@value
let $z := $b/attribute[@name='cust zip code']/@value
where $a/attribute/@name='cust city'
and $a/attribute/@value='Lyon'
and $x/dimension /@id=$a/@id
and $x/dimension/@id='customers'
group by(cust name,@cust zip code)
return name='cust name', aggregation(sum, quantity)
```

**Figure 2:** Sample decision-support XQuery

## 4. Join index structure

Building the indices cited in Section 2 on an XML warehouse causes a loss of information in decision-support query resolution. Indeed, clustering (1-index) or merging (data guide) identical labels causes the disappearance of the relationship between fact measures and dimensions. We illustrate this problem in the following example.

The *Facts.xml* document is composed of *Cell* elements. Each cell is identified by its dimension identifiers and one or more measures. Figure 3 shows the structure of the *Facts.xml* document and its corresponding 1-index (we selected the 1-index as an example). The 1-index represents cells linearly, i.e., all labels for the same source are represented by one label only. Hence, recovering a cell characterized by its measures and their dimension identifiers is impossible.

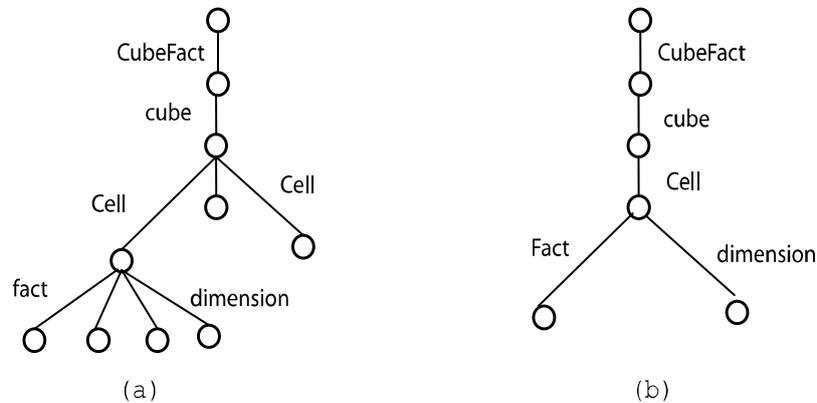

**Figure 3:** *Facts.xml* structure (a) and corresponding 1-index (b)

An index should be able to preserve the relationships between dimensions and fact measures. Thus, our index' structure is similar to that of the *Facts.xml* document, except for the attribute element.

XML indices usually summarize or reorganize the structure of the indexed XML documents into new XML documents that are then accessed instead of the original data. Our index structure is similar. It is stored in an XML document named *Index.xml*, whose structure is showed in Figure 4. Classically, graph labels starting with @ represent attributes. The others represent elements. Each Cell element is composed of dimensions and one or more facts. A *Fact* element has two attributes, *@id* and *@value*, which respectively represent measure names and values. Each *dimension* element is composed of two attributes: *@id*, which stores the dimension name, and *@node*, which stores the value of the dimension identifier. Moreover, the *dimension* element has children *attribute* elements. These elements are used to store the names and values of the attributes from each dimension. They are obtained from the *Dimensions.xml* document. An *attribute* element is composed of two attributes, *@name* and *@value*, which respectively store the name and value of each attribute.

Data migration from *Dimensions.xml* and *Facts.xml* to the index structure helps store facts, dimensions and their attributes in the same cell. This feature wholly eliminates join operations since all the information that is necessary for a join operation is stored in the same cell.

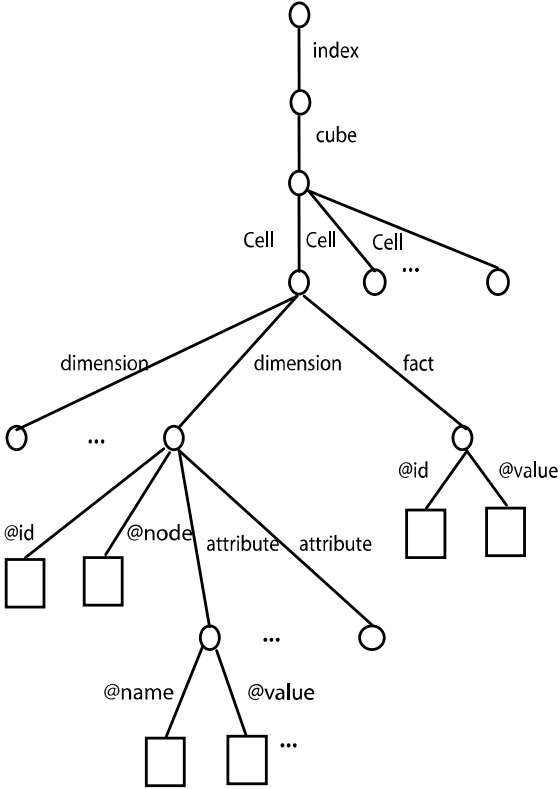

**Figure 4:** XML join index structure

Queries need to be rewritten to exploit our index. The rewriting process consists in preserving selection expressions and aggregation operations. We illustrate query execution by an example in Figure 5. More details are provided in Section 5.

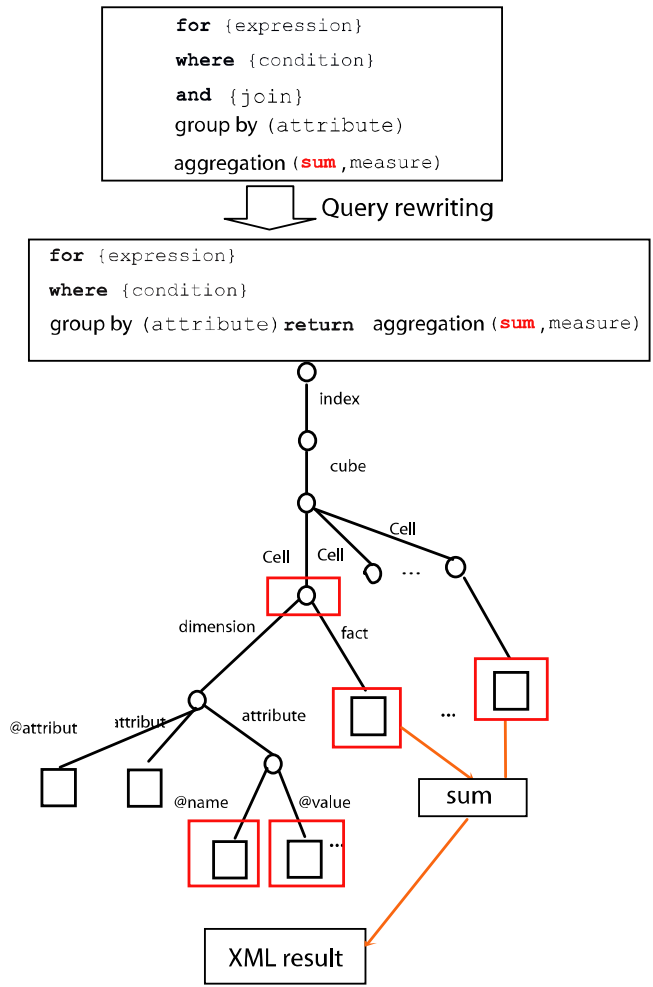

**Figure 5:** Query executions over our join index

## 5. Theoretical study

Queries defined over an XML warehouse modeled according to the XCube specification perform several join operations between facts stored in *Facts.xml* and dimensions from *Dimensions.xml*. Thus, they must satisfy the following constraints.

$$\text{document}(\textit{Facts.xml})/\text{CubeFact/cube/Cell/dimension}[@id = \\ \text{document}(\textit{Dimensions.xml})/\text{classification/Level/@id}] \\ \text{and} \\ \text{document}(\textit{Facts.xml})/\text{CubeFact/cube/Cell/dimension}/[@value = \\ \text{document}(\textit{Dimension.xml})/\text{classification/Level/node/@id}]$$

The first equality checks whether the dimension composing a cell (fact) is indeed the dimension expressed in the query. The second equality checks whether the node of a dimension (equivalent to a primary key) corresponds (can be joined) to the node, from the same dimension, defined in a cell (equivalent to a foreign key in the fact table).

Query execution without using our index may proceed as follows. For each dimension defined by *@node='name of dimension'*, the identifiers *@id* verifying the *Where* clause are searched for. The *Dimensions.xml* document is traversed in depth first, down to the *Level* node. Children nodes of the *Level* node are then traversed in breadth first until *@node* is equal to the dimension name.

This traversal cost is equal to the number of *Level* nodes in the *Dimensions.xml* document, denoted |dimension|. If several dimensions are defined in the query, all *Level* nodes are traversed for each dimension. Each node's child is traversed in depth first, until a list of *@id* attributes verifying the conditions *@name='name of the attribute'* and *@value='value of the attribute'* is found. The cost of this traversal is equal to the number of *attribute* children. Thus, dimension cost traversal is equal to $|a_i|*|d_i|$, where $|a_i|$ is the number of attributes in each dimension and $|d_i|$ the number of *node* elements, i.e., the number of children in each dimension.

To carry out a join between dimensions of the *Dimensions.xml* document and facts from the *Facts.xml* document, the *@id* values found when processing dimensions are searched for in the facts. The *Facts.xml* document is then traversed in depth first, down to the *Cell* level. Cells are then traversed in breadth first until dimensions whose child *@id* equals to *@node* in *Dimensions.xml* and *@node* equals to *@id* in *Dimensions.xml* are found. The traversal cost of the *Facts.xml* document is $|cell|$, where $|cell|$ is the number of cells. Finally, query execution cost without our index is:

$$E_{noindex} = ((|cell|*|\text{dimension}|)*(|\text{dimension}| + (|d_i|*|a_i|))).$$

Query execution when using our index may proceed as follows. For each dimension defined by *@node='name of dimension'*, the identifiers *@id* verifying the *Where* clause are searched for. The *Index.xml* document is traversed in depth first, down to the *Cell* level. The cost of this traversal is equal to the number of cells in the *Index.xml* document. *Cell dimension* children nodes are then traversed until the node whose *@id* value is equal to the dimension name in the query is reached. The cost of this traversal is equal to the number of *dimension* nodes in the *Index.xml* document, i.e., the number of dimensions in the warehouse schema. This cost is denoted |dimension|. The children of each found node are traversed in depth first, down to the *attribute* node verifying the conditions *@name='name of the attribute'* and *@value='value of the attribute'*. The cost of this traversal is equal to the number of *attribute* children, denoted $|a_i|$. Finally, query processing cost over our index structure is:

$$E_{index} = |cell|*(|\text{dimension}| + |a_i|).$$

Figure 6 shows the cost variation between $E_{noindex}$ and $E_{index}$ with respect to the number of cells (facts) from *Facts.xml*. These facts are described by five dimensions that are stored in the *Dimensions.xml* document. Table 1 displays the characteristics of these dimensions. We use a logarithmic scale on the Y axis to better visualize cost differences. Using our index induces a performance gain factor of 14,000 on an average.

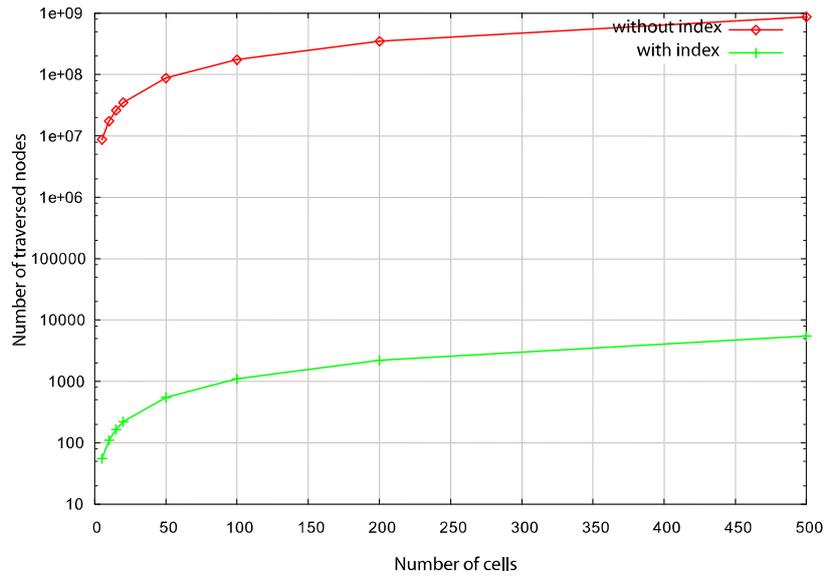

**Figure 6:** Theoretical results

| Facts | Number of cells |
|---|---|
| Sales | 16 260 336 |
| **Dimensions** | **Number of occurrences** |
| Customers | 50 000 |
| Products | 10 000 |
| Times | 1 461 |
| Promotions | 501 |
| Channels | 5 |
| **Documents** | **Size (MB)** |
| Facts.xml | 4.92 |
| Dimensions.xml | 3.77 |
| Schema.xml | 0.001 |

**Table 1:** Test data warehouse characteristics

# 6. Experiments

In order to validate our proposal experimentally, we generated an XML data warehouse modeled according to the XCube specification. Actual data have been transferred from an existing, sample relational data warehouse derived from an Oracle example[1].

This classical test data warehouse (Table 1), modeled as a star schema, is composed of *sale* facts characterized by the *amount* (of purchased products) and *quantity* (of purchased products) measures. The facts are stored in the *Facts.xml* document. They are described by five dimensions: *channels, promotions, customers, products* and *times* that are stored in the *Dimensions.xml* document.

---

[1] http://downloadwest.oracle.com/docs/cd/B10501 01/server.920/a96520/toc.htm

We implemented this data warehouse within two native XML DBMSs: eXist (Meier 2002) and X-Hive (Waldt 2005). Both these DBMSs allow the native storage of large documents and support the XQuery language. They also provide APIs (Application Programming Interfaces) for storing, querying, retrieving, transforming and publishing XML data.

We also implemented our XML data warehouse in a relational, XML-compatible DBMS: SQL Server (Rys 2004). SQL Server 2005 handles XML data through an XML type field. It integrates XQuery queries with the help of a function called query that is embedded into SQL Select clauses (Figure 7).

```
Select XML-DOC.Query('for $a in  //dimensionData/classification/Level[@node='customers']/node,
                     where $a/attribute/@name="cust city"
                     and $a/attribute/@value="Lyon"
                     return name="cust name"')
From DIMENSION
```

**Figure 7:** SQL-XQuery sample query

Our experiments measure the execution time of the typical decision-support query from Figure 2 over our test data warehouse, with and without exploiting our join index. We also varied warehouse size. Note that, in SQL Server 2005, XML data are stored in a table field. SQL-XQuery queries must be processed for each record. This process does not allow joining XML data from different records. Hence, we only performed our experiment with our join index on SQL Server, and not with the original, multi-document warehouse. We ran our tests on a Pentium 2 GHz PC with 1 GB of main memory and an IDE hard drive. Also note that we do not consider index construction time here, since an XML index is actually a new warehouse structure that is built once and queried thereafter.

Figure 8 represents our test query execution time on the three DBMSs, with and without using our index. The X axis represents warehouse size and the Y axis the corresponding execution time. The Y axis is in logarithm scale to highlight the differences in execution costs.

Our experimental results show that using our index structure significantly improves response time. On an average, the gain factor is indeed 25,669 for eXist and 8,411 for X-Hive. Furthermore, the plots we obtain are very similar in trend to those from Figure 6, which tends to validate our theoretical study.

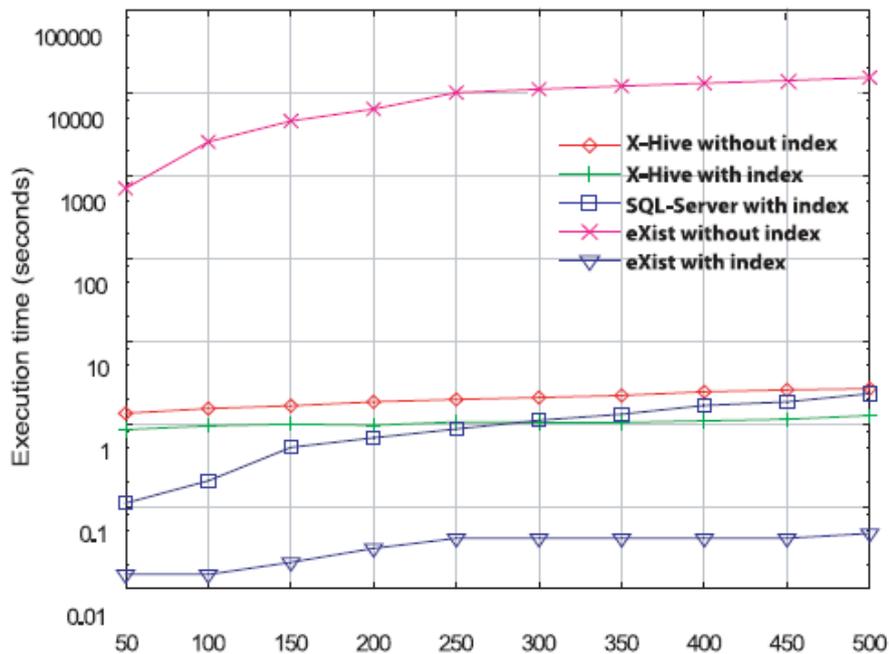

**Figure 8:** Experimental results

Though this is not plotted on Figure 8, we also pushed our "with/without index" tests further on the totality of the cells from the *Facts.xml* document. We achieved execution times of less than two seconds with our join index. Without index, X-Hive responded in about four minutes and eXist proved unable to answer in a reasonable time.

Finally, our experiments show that, properly indexed, native XML DBMSs can compete with, and even best relational DBMSs in terms of performance when XML documents are bulky. eXist running on our join index indeed outperforms SQL-Server by a 31.5 factor, on an average. This is because relational DBMS engines combine XQuery to SQL and must convert the result from relations to XML. XML native DBMSs, on the other hand, preserve the hierarchical structure of XML data, which allows path scans to be efficiently processed by XQuery engines. Our experiments also show that eXist's query engine performs better than X-hive when using simple path expressions. We think this is because eXist implements a specific numbering scheme that helps easily evaluate parent/child node relationships (Meier 2002).

## 7. Conclusion

In this paper, we presented a new join index that is specifically adapted to XML data warehouses. This data structure allows optimizing the access time to several XML documents by eliminating join costs, while preserving the information contained in the initial warehouse. Though we used an XCube-based reference data warehouse for illustration and validation purposes, our index structure can be built on any other XML warehouse model we are aware of.

To validate our proposal, we performed both a complexity study and experiments. We implemented our reference warehouse with two native XML DBMSs and one relational,

XML-compatible DBMS. Our tests showed that using our index structure significantly improves the response time of a typical decision-support query expressed in XQuery. Furthermore, they also demonstrate that native XML DBMSs can compete with and even best relational DBMSs.

This work also opens three broad axes of research perspectives. First, our indexing strategy could be better integrated into a host native XML DBMS. This would certainly help develop an incremental strategy for the maintenance of the join index data structure. Moreover, the mechanism for rewriting queries would also be more efficient if it was part of the system.

It is also crucial to carry on adapting or developing highly efficient optimization techniques in native XML DBMSs. XML views are getting more and more efficient, but there is still room for improvement, e.g., when generating and refreshing materialized views (Mahboubi et al. 2006). Partitioning an XML warehouse could also be envisaged.

Eventually, decision-support queries bear specific needs in terms of operators. For instance, we had to extend XQuery to allow multiple *group by* clauses to implement our decision-oriented workload. Similar extensions do exist already (Beyer et al. 2005), but it could be interesting to further extend XQuery to support OLAP operators such as *cube*, *rollup* or *drill-down*.